# Possible Josephson-like Behavior of the YBa$_2$Cu$_3$O$_{7-x}$ Single Crystal Twin Boundaries in Low Magnetic Fields

V.P.Timofeev[a] and A.V.Bondarenko[b]

[a]B.Verkin Institute for Low Temperature Physics & Engineering National Academy of Sciences of Ukraine, 47 Lenin Ave., 61103 Kharkov, Ukraine
[b]Physical Department, Kharkov State University, 4 Svobody Sq., 61077 Kharkov, Ukraine

**Abstract**
The experimental results on the magnetic response of YBCO single crystals with unidirectional twin boundary planes in low magnetic fields (0.01 ÷ 1 Oe) are discussed. The observed non-monotone temperature dependence of magnetization is interpreted within a plausible model of a system of the Josephson weak links in the twin boundary planes and on the basis of the order parameter anisotropy.

Studies of high-$T_c$ superconductors (HTSC) by magnetic methods are currently important both for understanding the nature of the high-$T_c$ superconductivity and for practical applications. The advantage of these methods is their unique sensitivity in squid-based magnetometry and the absence of contacts which can affect the surface structure of the object. Although HTSC cuprate was thoroughly studied theoretically and experimentally [1,2], there are still many open questions, for instance, about the magnetic flux dynamics, the origin of spontaneous current in a weak magnetic field and at close to critical temperatures [3]. These questions are important for development of LN$_2$-cooled high-sensitivity HTSC squids when it is necessary to reduce the intrinsic noises of magnetic sensors and to improve the sensitivity of the equipment.

The goal of the paper is to interpret the results on the non-monotone temperature dependence of the YBa$_2$Cu$_3$O$_{7-x}$ (YBCO) single crystal magnetization in very weak magnetic fields (0.01-0.8 Oe). The dependence is attributed to a possible thermoactivated transformation in the system of the Josephson weak links at the unidirectional twin boundaries in the region of the superconducting phase transition. The objects were pure *c*-oriented single crystal samples, grown by the solution-melted technology in the presence of a weak longitudinal gradient of the temperature [4]. After the annealing in oxygen, the single crystals had the typical critical temperature $T_c$ =93 K, measured by the resistive method in a "zero" magnetic field. The superconducting transition width (about 0.3 K) points to the high quality of the samples. The annealing at 400 °C required for optimization of doping leads to transformation of the tetragonal crystal structure into the orthorhombic one and, as a result, to the formation of the twin boundary planes. To investigate the influence of these plane defects on the pinning processes, we have selected YBCO single crystals with sizes close to 1×1×0,02 mm which had unidirectional twin boundaries oriented parallel to the *c*-axis of the crystal.

The temperature dependence of magnetization in low fields in the phase transition region was studied with the help of a helium cooled squid-based magnetic susceptometer (the "dBz/dZ" -type). The standard DC technique of magnetization measurement was used to register the RF squid response when moving the sample along the antenna axis in a uniform magnetic field of a solenoid [5]. The residual magnetic field of the Earth in the experimental chamber of intermediate temperature with the sample did not exceed 0.5 mOe. This was achieved by using multilayer mu-metal shields. This allowed cooling the sample and transferring it to the superconducting state by the ZFCW (zero field cooled warming) method.

The typical temperature dependence of the susceptometer output voltage at rising temperature of one of the samples is shown in Fig.1. The conversion ratio of the magnetic moment into the voltage response of the experimental setup is equal to $5\times10^{-10}$ A m$^2$/V (in the operating range of squid sensitivity ±10 $\Phi_0$, where $\Phi_0$ is the magnetic flux quantum). The magnetic field of the solenoid was directed along the *c*-axis of the single crystal and was equal to 15.5 А/м (≈0.2 Oe). With this orientation the magnetic field is parallel to the twin boundaries and the Abrikosov vortices are pinned most effectively. Similar dependence of good reproducibility was obtained for a number of other YBCO single crystals of the optimum doping level. It is seen in Fig.1 that in contrast to the resistive measurements for YBCO single crystals [4], the curve for superconducting phase transition is nonmonotonic and occupies the significant interval of temperatures $\Delta T = 3 - 5$ K. The dependence has a smoothed step which cannot be attributed to melting of the vortex lattices [1] taking into account very low magnetic fields in the experiment.

In this connection several important points should be indicated. While studying the magnetic susceptibility of polycrystalline YBCO and RBCO(R=Nd, Eu, Gd) cuprates, the authors of [6] found, that for some samples, annealed in oxygen, the ZFC phase transition curves have a smoothed step-like part (kink behavior). This type of the temperature dependence of susceptibility is typical for the glass – fluid transition in the vortex substance which occurs at $T_g<T_c$ in strong magnetic field, $H>H_{c1}$. Since in the experiment described the magnetic fields were low (H = 10-30 Oe), the observed step-like type of the phase features was compared with transformation of the Josephson weal links predicted theoretically [7] at low magnetic fields. The superconducting granules of a cuprate sample are bound by random barriers of different transparencies for Cooper pairs, and form statistical portioned networks for currents. The parameters of this Josephson medium depend on preparation technology of HTSC samples (pressures, degree of oxygen saturation, etc.), the sample temperature and external magnetic field. By analogy with spin glass, this state is sometimes called "orbital glass", implying that the orbital moments are associated with circular currents (eddy or spontaneous supercurrent). The latter can exist under certain condition, for instance, in the presence of an odd number of Josephson $\pi$-contacts in random current loops. The temperature of this transition $T_g$ shifts towards lower values as applied magnetic fields increase, and the step disappears in high magnetic fields where susceptibility has the typical diamagnetic behavior.

Similar effects were observed in our experiments on the YBCO single crystals. Fig.2. shows the magnetization normalized to its maximum value for one of the single crystals in three different magnetic fields H ∥ $c$, equal to 8 А/м ($\approx$0.1 Oe), 15.5 А/м ($\approx$0.2 Oe) and 65.9 А/м ($\approx$0.83 Oe). It is seen that the step shifts towards lower temperatures and becomes smoothed when the magnetic field increases.

Granular superconductors have complicated microstructures due to the anisotropy of the grains and are more favorable for the existence of orbital glass. The Abrikosov vertices are capable of penetrating into the grains along the $a$-$b$ planes. The supercurrent can flow through the loops connecting one or several grains. The similar characteristics were observed in high-quality YBCO films in higher magnetic fields (B = 0,3 T) [8]. This behavior was interpreted in the framework of a model based on a 1D array of Josephson junctions with statistically distributed lengths and field-dependent barrier thicknesses. The transport current there was interpreted as the current of tunneling through this network modulated by defects. At high temperature many of these channels break into separate superconducting granules, as a result, the critical current decreases.

The results of our experiments on YBCO single crystals suggest that twin boundary planes provide the condition for the formation of the similar Josephson networks with randomly distributed parameters. The twin boundaries include the $CuO_x$ layers, containing oxygen vacancies along these planes which produce local strong influence on the suppression of the superconducting order parameter. The suppression results in reduction of the energy of trapped vortices. The vortex density is therefore higher in the twin boundaries than on the rest of the crystal. The experiments based on Bitter-pattern technique [9] have shown that in magnetic fields of 20 - 40 Oe the vortex density in twins is about two times higher than in the crystal bulk. Considering that in field of 0.2 Oe the vortex spacing $a_0 = (\Phi_0/B)^{1/2} \approx 10^4$ nm exceeds the twin boundary spacing $d \approx 10^3$ nm and is comparable with the magnetic field penetration depth $\lambda \approx 10^4$ nm in this temperature interval, we can expect that all the vertices are localized on twins since vortex interaction decreases exponentially with increasing of the inter-vortex distance.

The superconducting surface regions, split up by twin boundaries, form a chaotic network for circulating eddy currents, which determine the resulting magnetization of the sample. When the temperature is increased, most of the weak Josephson junctions break, some of persistent supercurrents are redistributed and damped due to vortices creep. The twin boundary planes, being efficient centers of pinning, can also act as directions for relieved thermoactivated entry of the magnetic fluxes along these planes, the region of the possible generation of the spontaneous current [2]. To estimate the contribution of this mechanism to the behavior of the YBCO single crystal magnetization authors are planning to continue the study on the samples with crossed twin boundaries, and on untwined ones.

The full text of the paper will be published in Low Temperature Journal (Kharkov, 2004).

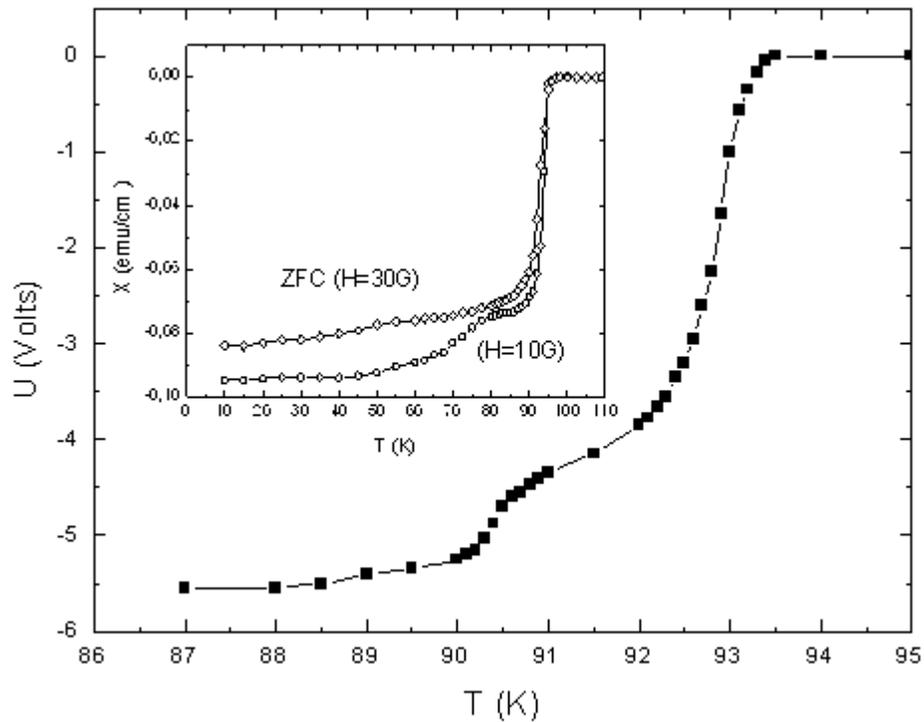

Fig.1. Susceptometer output signal at increasing temperature of the sample in the interval of the superconducting phase transition. Inset: the temperature dependence of susceptibility of polycrystalline YBCO from [6].

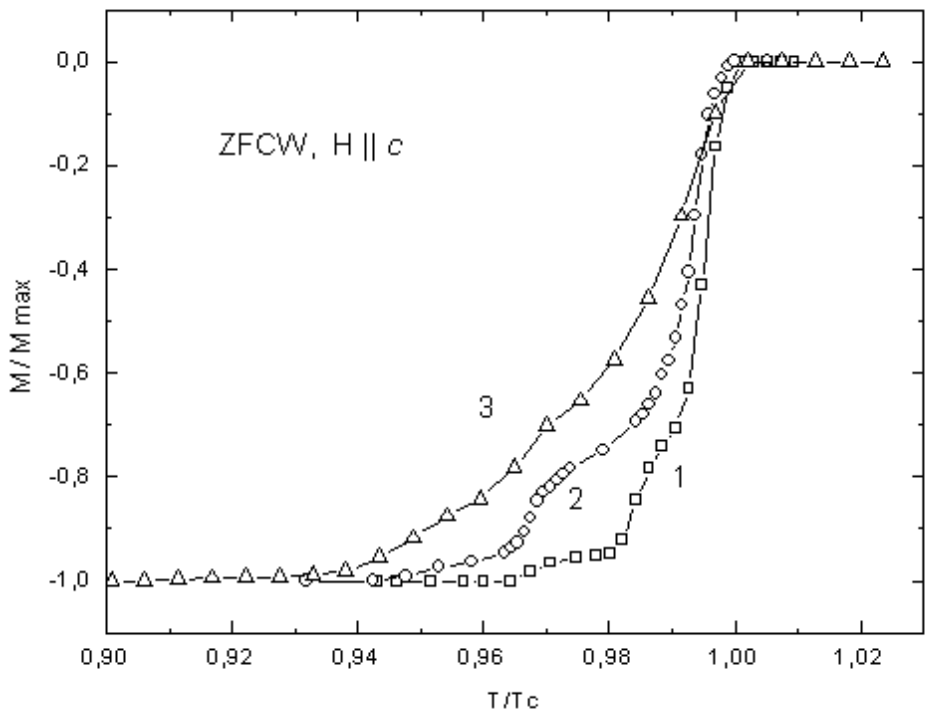

Fig.2. Temperature dependence of the normalized magnetization for a single crystal in different magnetic fields (*1* – 8 A/m; *2* – 15.5 A/m; *3* – 65.9 A/m).